\documentclass[11pt]{article}
\advance \textheight by \topskip
\newtheorem{theo}{Theorem}[section]
\newtheorem{ex}[theo]{Example}
\newtheorem{prop}[theo]{Proposition}

\newtheorem{defi}[theo]{Definition}

\newtheorem{rema}[theo]{Remark}

\usepackage{amscd}
\usepackage{amssymb}
\usepackage{amsmath}
\usepackage{eucal}
\usepackage{times}
 
\newcommand{\nc}{\newcommand}
\nc{\beq}{\begin{equation}} 
\nc{\eeq}{\end{equation}}
\nc{\beqa}{\begin{eqnarray}}
\nc{\eeqa}{\end{eqnarray}}

\def\cc#1{\kern .7em\hfill #1 \hfill\kern .7em}

\def\bb{\mathbb}
\begin{document}
\title{Finite-dimensional  Lie algebras of order $F$}
\author{M. Rausch de Traubenberg 
\thanks{rausch@lpt1.u-strasbg.fr}$\,\,$${}^{a,b}$ and
M.J. Slupinski \thanks{slupins@irmasrv2.u-strasbg.fr}$\,\,$${}^{c}$\\
\\
{\small ${}^{a}${\it Laboratoire de Physique Th\'eorique, Universit\'e 
Louis Pasteur}}\\
{\small  {\it 3 rue de
l'Universit\'e, 67084 Strasbourg, France}}\\
{\small ${}^{b}${\it Laboratoire de Physique
Math\'ematique et Th\'eorique,
Universit\'e Montpellier II,}}\\
{\small {\it Place E. Bataillon,  34095 Montpellier,
France}}\\
{\small ${}^c${\it Institut de Recherches en Math\'ematique Avanc\'ee}}\\
{\small { \it Universit\'e Louis-Pasteur, and CNRS}}\\
{\small {\it 7 rue R. Descartes, 67084 Strasbourg Cedex, France}}}
\date{\today}

%\begin{document}
\baselineskip=15pt
\maketitle

%\begin{document}

\begin{abstract}
$F-$Lie algebras are natural generalisations of Lie algebras ($F=1$)
and Lie superalgebras ($F=2$). When   $F>2$ not many finite-dimensional
examples are known. In this paper we construct
finite-dimensional $F-$Lie algebras  $F>2$ by an inductive process
 starting  from Lie algebras and Lie 
superalgebras. Matrix realisations of $F-$Lie algebras constructed
in this way from $\mathfrak{su}(n), \mathfrak{sp}(2n)$  
$\mathfrak{so}(n)$ and    $\mathfrak{sl}(n|m)$, 
$\mathfrak{osp}(2|m)$ are given.
 We obtain  non-trivial extensions of the Poincar\'e
algebra by In\"on\"u-Wigner contraction of certain $F-$Lie algebras with
$F>2$.

\end{abstract}

\section{Introduction}
\renewcommand{\theequation}{1.\arabic{equation}}   
\setcounter{equation}{0}   
The classification of algebraic objects satisfying certain axioms
may be considered a fundamental objective on  purely mathematical grounds. If
in addition, these objects turn out to to be relevant  for the
description of the possible symmetries of a physical system, such a
classification takes on a whole new meaning. 
The main question is, of course, what are the mathematical structures 
which are useful in describing the laws of physics.
Simple complex finite-dimensional Lie algebras were classified
at the end of the 19th century by W. Killing and E. Cartan well before 
any physical applications  were known.
Since then, Lie algebras have become essential for the description of
space-time symmetries and fundamental interactions.
On the other hand, it was  the discovery of supersymmetry in relativistic 
quantum field theory or as a possible extension of Poincar\'e invariance
\cite{susy} which gave rise to the concept of Lie superalgebras and
their subsequent classification  \cite{super,fk}. 

It is  generally accepted that because of the theorems of Coleman 
$\&$ Mandula \cite{cm}
and  Haag, Lopuszanski $\&$ Sohnius \cite{hls},
one cannot go beyond supersymmetry.  
However, if one weakens
the hypotheses of these two theorems, one can imagine symmetries 
which go beyond supersymmetry  
\cite{ker, luis, para,
fsusy, fsusy1d, am, fr, qfsusy, hek, asm, prs, fsusy2d, ad, kk, fvir, fvir2, 
fsusy3d, fsusyh, flie, brs, poly, infty,flie2},
the idea being that then the generators of the Poincar\'e algebra can be
obtained as an  appropriate product of more than two 
fundamental additional symmetries.
These new generators are in a representation of the Lorentz algebra
which is neither bosonic nor fermionic. Two kinds of representations are
generally taken:  parafermionic representations \cite{paraferm}, 
or infinite-dimensional representations (Verma module) \cite{verma}. 

Fractional supersymmetry (FSUSY) \cite{fsusy, fsusy1d, am, fr, qfsusy, 
hek, asm, prs, fsusy2d, ad, kk, fvir, fvir2, fsusy3d, fsusyh, flie, brs,
 poly, infty,flie2} is among  the possible extensions of supersymmetry 
which have been studied in the literature. 
Basically, in such extensions,
 the generators of the Poincar\'e algebra
are obtained as   $F-$fold ($F \in \bb N^\star$) 
symmetric products of more fundamental
generators. A natural generalisation of Lie (super)algebras
which is relevant for the algebraic description of 
FSUSY was defined in \cite{flie, flie2} and called an $F-$Lie algebra.
An $F-$Lie algebra admits a $\bb Z_F-$gradation, the zero-graded part
being a Lie algebra. An $F-$fold symmetric product (playing the role of
the anticommutator in the case $F=2$)  expresses the zero graded  part
in terms of the non-zero graded part.

The purpose of this paper is to show how one can construct many examples of
finite-dimensional $F-$Lie algebras by an inductive process starting from
Lie algebras and Lie superalgebras.
Some preliminary results is this direction  were given in \cite{flie2}.
Two types of
finite-dimensional $F-$Lie algebras will be constructed.
The  first  family of examples, which we call trivial, 
are obtained by taking the direct sum of  a Lie (super)algebra  with the
trivial representation. 
The second family  is  more interesting: by  an inductive procedure
we show how one can give the underlying vector space of any
Lie algebra or any classical Lie superalgebra the structure of
an $F-$Lie algebra. This procedure involves Casimir operators in the case
of  Lie algebras and invariant symmetric forms on the odd part of the
algebra in the case of Lie superalgebras.

The paper is organized as follows.
In   section 2 we recall the  definition of an $F-$Lie algebra
and show how one can construct an $F-$Lie algebra of order 
$F_1+F_2$  from an $F-$Lie algebra of order $F_1 \ge 2$ and an invariant 
symmetric form of order $F_2$   on its non-zero graded part
({\it c.f.} Theorem \ref{tensor}). In section 3 we introduce the
notion of a graded $1-$Lie algebra in order to prove a version of 
theorem \ref{tensor}  when $F_1=1$ (theorem \ref{tensor-bis}), and
give some explicit examples of $F-$Lie algebras associated to
Lie algebras.
In section 4 we give explicit examples of $F-$Lie algebras associated
to Lie superalgebras. 
In section 5 we obtain FSUSY extensions of the Poincar\'e algebra
by In\"on\"u-Wigner contraction of certain $F-$Lie algebras constructed
in the two previous sections.
In section 6  we define a notion of simplicity for  $F-$Lie
algebras and give examples of simple and non-simple $F-$Lie algebras.
Finally, in section 7 we give  finite-dimensional 
matrix realisations  of the $F-$Lie algebras of section 4
induced from  $\mathfrak{sl}(m|n)$ and $\mathfrak{osp}(2|2n)$ and a 
quadratic form. Using finite-dimensional matrices,
we also show that  the underlying vector spaces 
of the graded $1-$Lie algebras $\mathfrak{su}(n) \oplus \mathfrak{su}(n),
\mathfrak{so}(n) \oplus \mathfrak{so}(n)$ and
$\mathfrak{sp}(2n) \oplus\mathfrak{sp}(2n)$ can be given  
$F-$Lie algebra structures which cannot be obtained by our inductive process.

\section{$F-$Lie algebras}
\renewcommand{\theequation}{2.\arabic{equation}} 
\setcounter{equation}{0}   
\subsection{Definition of $F-$Lie algebras}
In this section,  we recall briefly the definition of $F-$Lie
algebras given in \cite{flie, flie2}. Let $F$ be a positive integer and let 
$q=e^{\frac{ 2 \pi i}{F}}$. We consider   $S$ a complex vector space
and $\varepsilon$ an automorphism of $S$ satisfying $\varepsilon^F=1$.
We set ${\cal A}_k=S_{q^k}, 1 \le k \le F-1$ and ${\cal B}=S_1$ 
($S_{q^k}$ is the eigenspace corresponding to the eigenvalue $q^k$ of 
$\varepsilon$). Then we have $S= {\cal B }\oplus_{k=1}^{F-1} {\cal A}_k$.

\begin{defi} \label{f-lie}
$S= {\cal B} \oplus_{k=1}^{F-1} {\cal A}_k$ is called a (complex)  $F-$Lie 
algebra if it is
endowed with the following structure:

\begin{enumerate}
\item ${\cal B}$ is a (complex) Lie algebra and 
${\cal A}_k, 1 \le k \le F-1$ are 
representations
of ${\cal B}$. If  $[ , ]$ denotes the bracket on ${\cal B}$ 
and the action of ${\cal B}$ on  $S$ it is clear that 
$\forall b  \in {\cal B}, \forall s \in S, [\varepsilon(b),\varepsilon(s)]=
\varepsilon\left([b,s]\right)$.
\item There exist multilinear ${\cal B}-$equivariant maps 
$\left\{~~, \cdots,~~ \right\}: {\cal S}^F\left({\cal A}_k\right)
\rightarrow {\cal B}$, where  ${ \cal S}^F(D)$ denotes
the $F-$fold symmetric product
of $D$. It is easy to see that 
$\left\{\varepsilon(a_1), \cdots, \varepsilon(a_F)\right\}=
\varepsilon\left(\left\{a_1, \cdots, a_F\right\}\right),
\forall a_1, \cdots, a_F \in {\cal A}_k.$
\item
For $b_i \in {\cal B}$ and $a_j \in {\cal A}_k$ the following
``Jacobi identities'' hold:

\beqa
\label{eq:jac}
&&\left[\left[b_1,b_2\right],b_3\right] +
\left[\left[b_2,b_3\right],b_1\right] +
\left[\left[b_3,b_1\right],b_2\right] =0 \nonumber \\
&&\left[\left[b_1,b_2\right],a_3\right] +
\left[\left[b_2,a_3\right],b_1\right] +
\left[\left[a_3,b_1\right],b_2\right]  =0 \nonumber \\
&&\left[b,\left\{a_1,\dots,a_F\right\}\right] =
\left\{\left[b,a_1 \right],\dots,a_F\right\}  +
\dots +
\left\{a_1,\dots,\left[b,a_F\right] \right\}  \nonumber \\
&&\sum\limits_{i=1}^{F+1} \left[ a_i,\left\{a_1,\dots,
a_{i-1},
a_{i+1},\dots,a_{F+1}\right\} \right] =0.
\qquad \qquad \qquad   \qquad  \qquad  (J_4) \nonumber
\eeqa
\end{enumerate}

\end{defi}

\noindent
Note that the three first identities are automatic but the fourth, which we
will refer to as $J_4$, is an extra constraint. 

\begin{rema}
An $F-$Lie algebra is more than a Lie algebra  $\mathfrak{g}_0$,
a representation $\mathfrak{g}_1$ of $\mathfrak{g}_0$ and a
$\mathfrak{g}_0-$valued $\mathfrak{g}_0-$equivariant symmetric  form 
on $\mathfrak{g}_1$.
Indeed, although the three first Jacobi identities are manifest in
this situation, the 
fourth is not necessarily true. As an example, consider
$\mathfrak{g}_0 = \mathfrak{sl}(2, \mathbb C)$ and $\mathfrak{g}_1
={\cal S}_{2k+1}, \ \ k \in \mathbb N$ (the  irreducible representation
of dimension $2k+2$). From the decomposition
${\cal S}^2 \left({\cal S}_{2k+1} \right) =
{\cal S}_{4k+2} \oplus {\cal S}_{4k-2} \oplus \cdots \oplus {\cal S}_2$
one has an $\mathfrak{sl}(2, \mathbb C)-$equivariant mapping from
${\cal S}^2 \left({\cal S}_{2k+1} \right) \longrightarrow {\cal S}_2
 \longrightarrow \mathfrak{sl}(2, \mathbb C)$. But
$\mathfrak{g} = \mathfrak{sl}(2, \mathbb C) \oplus {\cal S}_{2k+1}$
is not a Lie superalgebra (the fourth Jacobi identity is not 
satisfied) except when $k =0$ where it reduces to $\mathfrak{osp}(1|2)$.
\end{rema} 

\begin{rema} \label{1-2-Lie}
A $1-$Lie  algebra  is  a Lie algebra, and 
a $2-$Lie algebra  is a Lie superalgebra. We will also
refer to these objects as $F-$Lie algebras of order one and two
respectively.
\end{rema}
 
\begin{rema} \label{irred-flie}
Notice that $\{a_1, \cdots, a_F \}$ is only defined if the $a_i$ are
in the {\it same} ${\cal A}_k$ and that $\forall k=1,\cdots, F-1$, 
the spaces $S_k={\cal B} 
\oplus {\cal A}_k$ are $F-$Lie
algebras.\\

\begin{center}
\noindent
{\bf From now on, we consider only $F-$Lie algebras $S={\cal B} \oplus {\cal
A}$ such that ${\cal A}$ is an eigenspace of $\varepsilon$.}
\end{center}
\end{rema} 

\begin{rema}\label{cartan-weil}
If ${\mathfrak h} \subset {\cal B}$ is a Cartan subalgebra and
$F_{\lambda_1}, \cdots, F_{\lambda_F} \in {\cal F}$ are 
respectively of weight
$\lambda_1, \cdots, \lambda_F$, then 
$\left\{F_{\lambda_1}, \cdots, F_{\lambda_F}\right\} \in {\cal B}$ is
of weight $\lambda_1 + \cdots + \lambda_F$. In particular, if 
$\lambda_1 + \cdots + \lambda_F \ne 0$ is not a root of ${\cal B}$ this 
bracket is zero.

\end{rema}

This structure can be seen as a possible generalisation of Lie
algebras ($F=1$) or Lie superalgebras ($F=2$) and can be compared, in
some sense, to the ternary algebras ($F=3$) considered in \cite{tern},
and  to the $n-$ary algebras ($F=n$)  introduced
in \cite{vk} but in a different context.
We have shown \cite{flie, flie2} that all examples of FSUSY considered in the
literature can be described within the framework of 
$F-$Lie algebras.

\subsection{An inductive construction of $F-$Lie algebras}

Let $\mathfrak{g}$ be a complex Lie algebra and let ${\mathfrak{ r}}, 
{\mathfrak{ r^\prime}}$
be representations of $\mathfrak{g}$ such that there is a 
$\mathfrak{g}-$equivariant map
$\mu_F: S^F(\mathfrak{{r}}) \rightarrow \mathfrak{{r^\prime}}$.  We set:

$$S={\cal B}\oplus {\cal A}_1 = (\mathfrak{g} \oplus {\mathfrak{ r^\prime}}) 
\oplus {\mathfrak{ r}}.$$

\noindent
Then, 
${\cal B}=\mathfrak{g} \oplus {\mathfrak{ r^\prime}}$ is a  Lie algebra
as the semi-direct product of $\mathfrak{g}$ and $\mathfrak{{r^\prime}}$
(the latter with the trivial bracket). We can extend the action of  
$\mathfrak{g}$ on
${\mathfrak{ r}}$  to an action of ${\cal B}$ on  ${\mathfrak{ r}}$
by letting ${\mathfrak{ r^\prime}}$ act trivially on ${\mathfrak{ r}}$.
This defines the bracket $[~,~]$ on $S$. For the map $\left\{\cdots \right\}$
we take $\mu_F$. The first three Jacobi identities  
are clearly satisfied, and the fourth is also satisfied as
each term in the expression on the L.H.S. of $J_4$  vanishes. Hence $S$
is an $F-$Lie algebra.
There are two essentially opposite ways of giving explicit examples of
$F-$Lie algebras of this type. One can either start from
${\mathfrak g}$ and $\mathfrak{{r^\prime}}$ and extract an ``$F-$root'' of
$\mathfrak{{r^\prime}}$,
or one can decompose ${\cal S}^F({\mathfrak{ r}})$
into irreducible summands and project onto one of them  \cite{flie}.
The first approach is the   more difficult since,
in general, it  involves infinite-dimensional representation theory.
For example  if one starts with  $\mathfrak{{r^\prime}}=\mathfrak{D}_{\mu_1}$, 
the vector representation of $\mathfrak{so}(1,d-1)$
of highest weight $\mu_1$,
the representation $\mathfrak{{r}}=\mathfrak{D}_{\frac{\mu_1}{F}}$
of highest weight $\frac{\mu_1}{F}$,
 is not exponentialisable
(see {\it e.g.} \cite{kpr}) and does not define a representation of the Lie
group  $\overline{SO(1,d-1)}$, 
except when $d=3$ where such representations describe relativistic
anyons \cite{anyon}. The second approach on the other hand
will always give finite-dimensional
$F-$Lie algebras if one starts from finite-dimensional representations.

The following theorem gives an inductive procedure for constructing
finite-dimensional $F-$Lie algebras.

\begin{theo} \label{tensor}
Let $\mathfrak{g_0}$ be a Lie algebra and $\mathfrak{g}_1$ a representation
of $\mathfrak{g_0}$ such that    

(i) $S_1=\mathfrak{g_0} \oplus \mathfrak{g_1}$ is an $F-$Lie algebra of
order $F_1>1$; 

(ii) $\mathfrak{g_1}$ admits  a $\mathfrak{g_0}-$equivariant symmetric form
$\mu_2$ of order $F_2 \ge 1$.

\noindent
Then  $S=\mathfrak{g_0}  \oplus \mathfrak{g_1}$ admits 
an  $F-$Lie algebra structure of order $F_1+F_2$, which we call the
$F-$Lie algebra induced from $S_1$ and $\mu_2$.
\end{theo}

\noindent

\noindent
{\bf Proof:}
By hypothesis, there exist 
 $\mathfrak{g_0}-$equivariant maps
 $\mu_1:{\cal S}^{F_1}\left(\mathfrak{g_1}\right) \longrightarrow 
\mathfrak{g}_0$  and 
 $\mu_2:{\cal S}^{F_2}\left(\mathfrak{g_1}\right) \longrightarrow 
\mathbb C$. 
Now, consider $\mu:{\cal S}^{F_1+ F_2}\left(\mathfrak{g_1}\right) 
\longrightarrow \mathfrak{g}_0 \otimes\mathbb C \cong \mathfrak{g}_0$ defined 
by 

\begin{eqnarray}
\label{eq:tensor}
&& \hskip 4truecm
\mu(f_1,\cdots,f_{F_1+F_2})=  \\
&&\frac{1}{F_1 !}\frac{1}{F_2 !} \sum \limits_{\sigma \in S_{F_1 + F_2}}
\mu_1(f_{\sigma(1)},\cdots,f_{\sigma({f_{F_1}})}) \otimes 
\mu_2(f_{\sigma(f_{F_1+1})},\cdots,f_{\sigma(f_{F_1+F_2})}),
\nonumber
\end{eqnarray}

\noindent
where $f_1,\cdots,f_{F_1+F_2} \in \mathfrak{g_1}$ and 
$S_{F_1 + F_2}$ is the group of permutations on $F_1 + F_2$ elements.
By construction, this is a $\mathfrak{g_0}-$equivariant map
from ${\cal S}^{F_1+ F_2}\left(\mathfrak{g_1}\right) \longrightarrow
\mathfrak{g_0}$, 
thus the three first Jacobi identities are  satisfied.  The last Jacobi
identity $J_4$,  is more difficult to check and  is a consequence of
$J_4$ for the $F-$Lie algebra $S_1$ and  a factorisation property.
Indeed,  setting $F=F_1+F_2$, the identity $J_4$ for the terms in

$$
\sum\limits_{i=0}^{F} \left[ f_i,\mu\left(f_1,\dots,
f_{i-1},
f_{i+1},\dots,f_{F}\right) \right],
$$

\noindent
of the form 
$\mu_1(f_{\sigma(1)},\cdots,f_{\sigma({f_{F_1}})}) \otimes 
\mu_2(f_{\sigma(f_{F_1+1})},\cdots,f_{\sigma(f_{F_1+F_2})}$
with $\sigma \in S_{F_1+F_2 +1}$, reduces to

\begin{eqnarray}
\sum\limits_{i=0}^{F_1} 
 \left[ f_{\sigma(i)},\mu_1\left(f_{\sigma(1)},\cdots,
f_{\sigma(i-1)},f_{\sigma(i+1)},\cdots,f_{\sigma({f_{F_1}})}
\right) \right]
\otimes 
\mu_2(f_{\sigma(f_{F_1+1})},\cdots,f_{\sigma(f_{F_1+F_2})}=0,
\nonumber
\end{eqnarray} 

\noindent
using $\mu_2(f_{\sigma(f_{F_1+1})},\cdots,f_{\sigma(f_{F_1+F_2})}
 \in \mathbb C$.
\noindent
But the L.H.S.  vanishes by  $J_4$  for
the $F-$Lie algebra $S_1$. 
A similar argument works for the other terms and hence $J_4$
is satisfied and $S$ is an $F-$Lie algebra of order $F_1 + F_2$.
{\bf QED}

\begin{rema}
Theorem \ref{tensor} is equivalent to the fact that
the product of two $\mathfrak{g_0}-$equivariant  symmetric forms
satisfying $J_4$ also satisfies $J_4$ if one of them is scalar-valued.
\end{rema}

%\begin{cor}
%Let $\mathfrak{g_0}$ be a Lie algebra and $\mathfrak{g}_1$ a representation
%of $\mathfrak{g_0}$ such that    

%(i) $S_1=\mathfrak{g_0} \oplus \mathfrak{g_1}$ is an $F-$Lie algebra of
%order $F_1>1$; 

%(ii) $\mathfrak{g_1}$ admits  a $\mathfrak{g_0}-$equivariant
% antisymmetric form
%of order $F_2>1$.

%\noindent
%Then  $S=\big(\mathfrak{g_0} \big) \oplus \left(\mathbb C^{F_2} \otimes
%\mathfrak{g_1}\right)$ admits 
%a structure of $F-$Lie algebra of order $F_1+F_2$. 
%\end{cor}

\section{Finite-dimensional $F-$Lie algebras associated to Lie algebras}
\renewcommand{\theequation}{3.\arabic{equation}} 
\setcounter{equation}{0}  

In this section we  first introduce the notion of a graded $1-$ Lie 
algebra in order to have a version of 
\ref{tensor} when $F_1=1$.

\subsection{Graded $1-$Lie algebras}

\begin{defi}\label{1-lie}
A graded $1-$Lie algebra is
a $\mathbb Z_2-$graded vector space
$S={\cal B} \oplus {\cal F}$ such that:

\begin{enumerate}
\item ${\cal B}$ is a Lie algebra;
\item ${\cal F}$ is a representation of ${\cal B}$;
\item there is a ${\cal B}-$equivariant map $\mu : {\cal F} \to {\cal B}$;
\item $[\mu(f_1),f_2]+ [\mu(f_2),f_1]=0$,  $\forall f_1, f_2 \in {\cal F}$.
\end{enumerate}
\end{defi}

\begin{ex} \label{graded}
Let $\mathfrak{g}$ be a Lie algebra.   Set  ${\cal B} = \mathfrak{g}$,
${\cal F} = \mathrm {ad } \  \mathfrak{g}$ and 
$S= {\cal B} \oplus {\cal F}$. If 
$\mu : {\cal F} \to {\cal B}$ is the identity then $(S,\mu)$ is a graded
$1-$Lie algebra.
\end{ex}

\begin{rema} \label{natural-lie}
A graded $1-$Lie algebra is not {\it a priori} a Lie algebra but it easy
to see that, in fact,  it has a natural graded Lie algebra structure.
\end{rema}

\begin{rema}
${\mathrm Ker} \mu$ is a ${\cal B}-$invariant subspace of ${\cal F}$ and
${\mathrm Im} \mu$ is a ${\cal B}-$invariant subspace of ${\cal B}$.
In particular, if ${\cal B}$ is simple, ${\cal F}$  irreducible and
$\mu$  non-trivial, then $\mu$ defines a ${\cal B}-$equivariant
 isomorphism between ${\cal F}$
and ${\cal B}$.
\end{rema}

A graded $1-$Lie algebra is a graded Lie algebra in 
the usual sense. In general, however,  a graded Lie algebra is not 
a graded $1-$Lie algebra since there is  no  preferred map from the odd
to the even part.

\begin{prop} 
Let $\mathfrak{g}=\mathfrak{g}_+ \oplus \mathfrak{g}_- $ be a graded
Lie algebra,  and let $\mu : \mathfrak{g} 
\to \mathfrak{g} $ be an odd $\mathfrak{g}_+-$equivariant  map
of $\mathfrak{g}$ such that $\mu $ is injective on 
 $[\mathfrak{g}_+, \mathfrak{g}_-]$. 
Then  $(\mathfrak{g}, \mu)$ is a graded $1-$Lie
algebra.
\end{prop}

\noindent
{\bf Proof:}
One only has to check \ref{1-lie}(4).
One has  $\forall f_1,f_2 \in\mathfrak{g}_-$,
$\mu\left([\mu(f_1),f_2] + [\mu(f_2),f_1]\right)=
[\mu(f_1),\mu(f_2)] + [\mu(f_2),\mu(f_1)]=0$.
Since $\mu$ is injective  on $[\mathfrak{g}_+, \mathfrak{g}_-]$
this implies \ref{1-lie}(4).
{\bf QED}

\begin{theo}[2.6-bis] \label{tensor-bis}
Let $\mathfrak{g_0}$ be a Lie algebra and $\mathfrak{g}_1$ a representation
of $\mathfrak{g_0}$ such that    

(i) $S_1=\mathfrak{g_0} \oplus \mathfrak{g_1}$ is an graded $1-$Lie algebra; 

(ii) $\mathfrak{g_1}$ admits  a $\mathfrak{g_0}-$equivariant symmetric 
$\mu_2$ form of order $F_2 \ge 1$.

\noindent
Then  $S=\mathfrak{g_0}  \oplus \mathfrak{g_1}$ admits 
an $F-$Lie algebra structure  of order $1+F_2$ which we call the $F-$Lie
algebra induced from $S_1$ and $\mu_2$. 

\end{theo}

\noindent
{\bf Proof:}
Analogous to \ref{tensor}.
{\bf QED}

\subsection{Trivial and induced $F-$Lie algebras} 

Consider the graded $1-$Lie algebra
$S=\mathfrak{g}_0 \oplus \mathfrak{g}_1$
where $\mathfrak{g}_0$ is a  
Lie algebra,  $\mathfrak{g}_1$ is the adjoint representation of 
$\mathfrak{g}_0$ and $\mu : \mathfrak{g}_1 \to \mathfrak{g}_0$ is the
identity. Let $J_1,\cdots, J_{\mathrm{dim}\mathfrak{g}_0}$ be
a basis of $\mathfrak{g}_0$, and $ A_1,\cdots,
 A_{\mathrm{dim}\mathfrak{g}_0}$ the corresponding basis of $\mathfrak{g}_1$.
The graded $1-$Lie algebra structure on $S$ is then:

\begin{eqnarray}
\label{eq:1-lie}
\left[J_a, J_b \right] = f_{ab}^{\ \ \ c} J_c, \qquad
 \left[J_a, A_b \right] = f_{ab}^{\ \ \ c} A_c, \qquad
\mu(A_a)= J_a,
\end{eqnarray}

\noindent
where $f_{ab}^{\ \ \ c} $ are the structure constant of $\mathfrak{g}_0$,
Two types of $F-$Lie algebras associated to $S$ will be defined.

The first type of $F-$Lie algebras associated to $S$, will be called trivial
and are constructed as follows:

\begin{theo}\label{lie-trivial}
Let $\mathfrak{g}_0$ be a Lie algebra and let $F \ge 1$ 
be an integer.
Then $S=\mathfrak{g}_0 \oplus \left( \mathfrak{g}_1 \oplus 
\mathbb C \right)$ can be given the structure of an $F-$Lie algebra
(graded $1-$Lie algebra if $F=1$)
where $\mathfrak{g}_1 $ is the adjoint representation of 
$\mathfrak{g}_0$ and  
 $ \mathbb C$ is the trivial representation.

\end{theo}

\noindent
{\bf Proof:}.
The map $\mu: {\cal S}^F(\mathfrak{g}_1 \oplus {\mathbb C}) 
\longrightarrow \mathfrak{g}_0$
is given by projection on $\mathfrak{g}_1$  in the decomposition
${\cal S}^F(\mathfrak{g}_1 \oplus {\mathbb C })=
{\cal S}^F(\mathfrak{g}_1) \oplus
{\cal S}^{F-1} (\mathfrak{g}_1) \oplus \cdots \oplus 
{\cal S}^{2} (\mathfrak{g}_1) 
\oplus \mathfrak{g}_1 
\oplus \mathbb C$, followed by the identification
of $\mathfrak{g}_1$ with $\mathfrak{g}_0$.  

\noindent
With the notations of (\ref{eq:1-lie})
the  brackets are:

\beqa
\label{eq:t-lie}
\left\{\lambda, \cdots, \lambda \right\}&=&0 \nonumber \\
\left\{\lambda,\cdots,\lambda, A_a,  \right\}&=& J_a \nonumber  \\
&\vdots& \\
\left\{\lambda,\cdots,\lambda, A_{a_1}, \cdots, A_{a_k} \right\}&=&0,
\qquad 1 < k \le F
\nonumber \\
&\vdots& \nonumber \\
\left\{A_{a_1},\cdots, A_{a_F} \right\}&=&0. \nonumber
\eeqa

\noindent
with $A_a,  \in \mathfrak{g}_1, \lambda \in \mathbb C$,
 $J_a \in \mathfrak{g}_0$.

It is easy to check that the four Jacobi identities are satisfied.
{\bf QED}\\

The second type of $F-$Lie algebras associated to $S$ are those induced
from $S$ and Casimir operators of $\mathfrak{g}_0$ (see \ref{tensor-bis}). 
It is well known that  the invariant tensors  on $\mathfrak{g}_0^\star$
are generated by  primitive invariant tensors which are either fully symmetric
or fully antisymmetric \cite{ec}.
By duality, symmetric invariant tensors are related to the Casimir operators 
of  $\mathfrak{g}_0$,
and it is well known that for a rank $r$ Lie algebra one can find
$r$ independent primitive Casimir operators.

\begin{theo} \label{casimir} 
Let $\mathfrak{g}_0$ be   a simple (complex) Lie algebra  and 
$\mathfrak{g}_1$ be the adjoint representation of $\mathfrak{g}_0$.
Then a Casimir operator of $\mathfrak{g}_0$ of order $m$ induces the
structure of an $F-$Lie algebra of order $m+1$ on 
$S_{m+1}= \mathfrak{g}_0 \oplus \mathfrak{g}_1$.
\end{theo}

\noindent
{\bf Proof:}
By example \ref{graded} $\mathfrak{g}_0 \oplus \mathfrak{g}_1$ is
a graded $1-$Lie algebra and the result follows from \ref{tensor-bis}.
{\bf QED}

\begin{rema} \label{induced-graded}
One can give explicit formulae for the bracket of these $F-$Lie
algebras as follows.
Let 
$J_a, a=1,\cdots, \mathrm{dim}(\mathfrak{g}_0)$ 
 and  let $A_a, a=1,\cdots, \mathrm{dim}(\mathfrak{g}_0) $
be bases as   at the beginning of this section.
Let $h_{a_1 \cdots a_{m}}$ be a  Casimir operator
of order $m$ (for $m=2$, the Killing form
 $g_{ab}=\mathrm{Tr}(A_a A_b)$ is a primitive Casimir of
order two). 
Then, the $F-$bracket of the  $F-$Lie algebra is 

\begin{eqnarray}
\label{eq:mi-lie}
\left\{A_{a_1}, A_{a_2}, \cdots, A_{a_{m+1}} \right\} =
\sum \limits_{\ell =1}^{m+1}
h_{a_1 \cdots a_{\ell-1} a_{\ell +1} \cdots a_{m+1}} J_{a_\ell}
\end{eqnarray}

\noindent
For the Killing form  this gives

\begin{eqnarray}
\label{eq:3-lie}
\left\{A_a, A_b, A_c \right\} =
g_{ab} J_c + g_{ac} J_b + g_{bc} J_a.
\end{eqnarray}         

\end{rema}

If $\mathfrak{g}_0= \mathfrak{sl}(2)$, the $F-$Lie algebra of
order three induced from the Killing form is the $F-$Lie algebra
of \cite{ayu}. 

\section{Finite-dimensional $F-$Lie algebras associated to Lie 
superalgebras}
\renewcommand{\theequation}{4.\arabic{equation}}
\setcounter{equation}{0}      
In this section we will consider some $F-$Lie
algebras which can be associated to Lie superalgebras using
Theorem \ref{tensor}.
\subsection{Lie superalgebras}
We first recall  some basic  results
on simple complex  Lie superalgebras (for more details  see \cite{r,fss}).
Simple Lie superalgebras  can be divided into  two types:
classical and the Cartan-type.  Classical
Lie superalgebras can be further divided into two families:
 basic and strange. A basic Lie superalgebra 
$\mathfrak{g}= \mathfrak{g}_0 \oplus \mathfrak{g}_1$ is said to be 
 respectively of type I or type II depending on whether the 
$\mathfrak{g}_0-$module
$\mathfrak{g}_1$ is respectively reducible or irreducible. Here is
the complete list of simple classical Lie superalgebras
\cite{super, fk}. In  the statement of 1(i)  
the symbol $(\overline{\mathbf { m+1}}, {\mathbf {n+1}})^+
\oplus ({\mathbf { m+1}}, \overline{\mathbf {n+1}})^-$
denotes $\left({\mathbb C}^{m+1\ \star} \otimes {\mathbb C}^{n+1}
\otimes {\mathbb C} \right)
\oplus \left({\mathbb C}^{m+1\ } \otimes {\mathbb C}^{n+1 \star}
\otimes {\mathbb C}^\star \right)$, where 
${\mathbb C}^{m+1 }$ is the fundamental 
representation of $\mathfrak{sl}(m+1)$, ${\mathbb C}^{m+1\star}$ its
dual representation and ${\mathbb C}$ is the standard one dimensional 
representation of $\mathfrak{gl}(1)$. In the rest of
the theorem we use  analogous notation.

\begin{theo} \label{super}
Let $\mathfrak{g}= \mathfrak{g}_0 \oplus \mathfrak{g}_1$ be a classical 
simple complex Lie superalgebra. Then $\mathfrak{g}$ is isomorphic to one
of the following:

\begin{enumerate}  
\item (Basic of type I) 

(i) $A(m,n)$: $m > n  \ge  0,  
\mathfrak{g}_0= \mathfrak{sl}(m+1) \oplus \mathfrak{sl}(n+1) 
\oplus  \mathfrak{gl}(1),  
\mathfrak{g}_1=(\overline{\mathbf { m+1}}, {\mathbf {n+1}})^+
\oplus (\mathbf{m+1}, \overline {{\mathbf {n+1}}})^-$

(ii)  $A(n,n): n \ge 1, \mathfrak{g}_0= \mathfrak{sl}(n+1) 
\oplus \mathfrak{sl}(n+1),  
\mathfrak{g}_1=(\overline{{\mathbf  {n+1}}}, \mathbf{n+1}) \oplus
({\mathbf {n+1}}, \overline {{\mathbf {n+1}}})$;

(iii)  $C(n+1): n \ge 1, \mathfrak{g}_0=\mathfrak{sp}(2n) 
\oplus \mathfrak{gl}(1), 
\mathfrak{g}_1={\mathbf {2n}}^+ \oplus {\mathbf {2n}^-}$. 

\item  ( Basic of type II) 

(i) $B(m,n): m \ge 0, n \ge 1, \mathfrak{g}_0= 
\mathfrak{so}(2m+1) \oplus 
\mathfrak{sp}(2n), \mathfrak{g}_1=(\mathbf{2m+1}, \mathbf{2n})$;

(ii) $D(m,n): m \ge 2, n \ge 1, m \ne n+1, \mathfrak{g}_0= 
\mathfrak{so}(2m) \oplus 
\mathfrak{sp}(2n), \mathfrak{g}_1= (\mathbf{2m}, \mathbf{2n})$;

(iii) $D(n+1,n): \mathfrak{g}_0= \mathfrak{so}(2(n+1))
 \oplus 
\mathfrak{sp}(2n), \mathfrak{g}_1 =(\mathbf{2(n+1)}, \mathbf{2n})$;

(iv) $D(2,1; \alpha): \alpha \in \bb C-\left\{0,-1\right\}, 
\mathfrak{g}_0= \mathfrak{sl}(2) \oplus \mathfrak{sl}(2) \oplus 
\mathfrak{sl}(2),
\mathfrak{g}_1=(\mathbf{2},\mathbf{2},\mathbf{2})$;

(v) for $F(4): \mathfrak{g}_0=\mathfrak{sl}(2) \oplus 
\mathfrak{so}(7), 
\mathfrak{g}_1=(\mathbf{2}, \mathbf{8})$;

(vi) for $G(3): \mathfrak{g}_0=\mathfrak{sl}(2) \oplus G_2, 
\mathfrak{g}_1 =(\mathbf{2}, \mathbf{7})$.

\item (Strange)

\noindent
 (i) $Q(n): n>1 \mathfrak{g}_0=\mathfrak{sl}(n),
\mathfrak{g}_1 =\mathrm{ad}(\mathfrak{sl}(n))$, with ad the adjoint
representation;

\noindent
 (ii) $P(n): n>1 \mathfrak{g}_0=\mathfrak{sl}(n),
\mathfrak{g}_1 = [2] \oplus [1^{n-2}]$, where $[2]$ denotes
${\cal S}^2\left({\mathbb C}^n\right)$ the two-fold symmetric
representation and $[1^{n-2}]$ denotes $\Lambda^{n-2}\left(\mathbb C^n\right)$
the $(n-2)-$fold antisymmetric representation.
\end{enumerate}
(The superscript in 1(i) and 1(iii) indicates the $\mathfrak{gl}(1)$
charge).
\end{theo}

\subsection{Symmetric invariant  forms} \label{sym-inv}
%Starting from the Lie superalgebra $\mathfrak{g}=\mathfrak{g}_0 \oplus
% \mathfrak{g}_1$, using theorem  \ref{tensor} one can construct $F-$Lie
%algebras from $\mathfrak{g}_0-$invariant symmetric forms on 
%$\mathfrak{g}_1$. 
By Theorem \ref{tensor} one can construct an $F-$Lie algebra from a Lie
superalgebra $\mathfrak{g}=\mathfrak{g}_0 \oplus \mathfrak{g}_1$ and a
$\mathfrak{g}_0-$invariant symmetric form on 
$\mathfrak{g}_1$.
In general determining  {\it all}
invariant symmetric forms on a given  representation of a given Lie
algebra is very difficult. However, for the Lie superalgebras given
in the above list  we will show how one can construct many invariant symmetric 
forms.  The key observation is that for each basic  Lie
superalgebra in the list, the odd part  $\mathfrak{g}_1$  is either a tensor
product (type II) or a sum of two dual tensor products (type I) as a 
representation of  $\mathfrak{g}_0$.
Thus, to find  $\mathfrak{g}_0-$invariant symmetric forms
on  $\mathfrak{g}_1$  one can use the
following well known isomorphisms of representations of
$GL(A) \times GL(B)$ \cite{fulton-harris}:

\beqa
\label{sum}
{\cal S}^p \left(A \oplus B \right)&=&
\sum \limits_{k=0}^p {\cal S}^k \left(A \right)\otimes 
{\cal S}^{p-k} \left(B\right)  \\
\label{Young}
{\cal S}^p \left(A \otimes B\right)&=& 
\sum \limits_{\Gamma} {\$}^{\Gamma} \left(A\right) \otimes
{\$}^{\Gamma} \left(B\right), 
\eeqa

\noindent
where the second  sum is taken over all Young diagrams $\Gamma$ of length $p$
and ${\$}^{\Gamma}\left(A\right)$ denotes the irreducible representation
of $GL(A)$ corresponding to the Young symmetriser of $\Gamma$.

\subsubsection*{Type I}
We consider  the Lie superalgebra $A(m,n)$. The case of the other
basic type I Lie superalgebras is similar. 
Then $\mathfrak{g}_0= \mathfrak{sl}(m+1) \oplus \mathfrak{sl}(n+1)
\oplus \mathfrak{gl}(1)$ and 
$\mathfrak{g}_1= \left({\mathbb C}^{m+1\ \star} \otimes {\mathbb C}^{n+1}
\otimes {\mathbb C} \right) \oplus
\left({\mathbb C}^{m+1 \star } \otimes {\mathbb C}^{n+1}
\otimes {\mathbb C} \right)^\star$. Using the formulae (\ref{sum}) and
(\ref{Young}), one sees that ${\cal S}^p({\mathfrak{g}}_1)$ is a direct
sum of terms of the form:

\beqa
{\$}^\Gamma \left({\mathbb C}^{m+1 \star}\right) \otimes 
{\$}^{\Gamma^\prime} \left({\mathbb C}^{m+1 }\right)\otimes 
{\$}^\Gamma \left({\mathbb C}^{n+1 }\right)\otimes 
{\$}^{\Gamma^\prime} \left({\mathbb C}^{n+1 \star }\right)\otimes 
{\mathbb C}^{|\Gamma|-|\Gamma^\prime|},
\eeqa

\noindent
where $|\Gamma|$ is the length of $\Gamma$ and 
$|\Gamma | + |\Gamma^\prime| =p$.
If this term contains the trivial representation then $n$ must be even and
$|\Gamma | =|\Gamma^\prime|$.
Furthermore the dimension of the vector space of $\mathfrak{g}_0$
invariants is then

\beqa
I_{\Gamma, \Gamma^\prime}=\mathrm {dim}\  \mathrm {Hom}_{\mathfrak{sl}(m+1)}
\left( {\$}^{\Gamma^\prime} \left({\mathbb C}^{m+1 }\right),
{\$}^{\Gamma} \left({\mathbb C}^{m+1 }\right)\right) \times
\mathrm {dim} \ \mathrm {Hom}_{\mathfrak{sl}(n+1)}
\left( {\$}^{\Gamma^\prime} \left({\mathbb C}^{n+1 }\right),
{\$}^{\Gamma} \left({\mathbb C}^{n+1 }\right)\right),
\eeqa

\noindent
where $\mathrm {Hom}_{\mathfrak{sl}(m+1)}$ denotes 
 homomorphisms which are 
$\mathfrak{sl}(n+1)$ equivariant. One can calculate the
dimensions of these spaces using well known results \cite{fulton-harris}. 
If $\Gamma= \Gamma^\prime$ then $I_{\Gamma, \Gamma^\prime} \ge 1$;
if $\Gamma=\Gamma^\prime$ and $|\Gamma|=|\Gamma^\prime|=1$ then
$I_{\Gamma, \Gamma^\prime} = 1$ and the invariant quadratic form corresponds
to the tautological metric on $\mathfrak{g}_1$. In \cite{flie2} 
$F-$Lie algebras were constructed using this symmetric form.

\subsubsection*{Type II}
All basic type II Lie superalgebras except (iv) have  $\mathfrak{g}_0=
\mathfrak{g}_0^\prime \oplus \mathfrak{g}_0^{\prime \prime}$
and $\mathfrak{g}_1= {\cal D}^\prime \otimes {\cal D}^{\prime \prime}$,
where  ${\cal D}^\prime$ and 
${\cal D}^{\prime \prime}$ are irreducible
self-dual representations of respectively
$\mathfrak{g}_0^\prime$ and  $\mathfrak{g}_0^{\prime \prime}$.
Therefore ${\cal S}^p\left(\mathfrak{g}_1 \right)$ is the direct sum
of terms of the form:

\beqa
\label{II}
{\$}^\Gamma\left({\cal D}^\prime\right) \otimes
{\$}^\Gamma\left({\cal D}^{\prime \prime}\right)
\eeqa

\noindent
where $|\Gamma|=p$. The dimension of the vector space of $\mathfrak{g}_0$
invariants is 
\beqa
I_{\Gamma}=\mathrm {dim} \
{\$}^\Gamma\left({\cal D}^\prime\right)^{\mathfrak{g}_0^\prime}
\times 
\mathrm {dim} \ {\$}^\Gamma\left({\cal D}^{\prime \prime}
\right)^{\mathfrak{g}_0^{\prime \prime}},
\eeqa

\noindent
where ${\$}^\Gamma\left({\cal D}^\prime\right)^{\mathfrak{g}_0^\prime}$
denotes the space of $\mathfrak{g}_0^\prime$ invariant vectors in
${\$}^\Gamma\left({\cal D}^\prime\right)$.
Although the respective factors in the product (\ref{II})
are irreducible for $GL({\cal D}^\prime)$ and
$GL({\cal D}^{\prime \prime})$, they may become reducible for 
$\mathfrak{g}_0^\prime$, $\mathfrak{g}_0^{\prime \prime}$. For example
the representations associated to the Young diagram
$\begin{tabular}{|c|c|}\hline
& \\ \hline
& \\
\cline{1-2}  
\end{tabular}$ are reducible for both 
$\mathfrak{g}_0^\prime = \mathfrak{so}(m)$ and 
$\mathfrak{g}_0^{\prime \prime} = \mathfrak{sp}(2n)$.

\subsubsection*{The strange superalgebra $Q(n)$}
Up to duality ${\cal S}^\star (\mathfrak{g}_1)$ (the symmetric
algebra on $\mathfrak{g}_1$) is generated by the Casimir operators
of $\mathfrak{sl}(n)$ (see section 3.).

\subsubsection*{The strange superalgebra $P(n)$}
In this case ${\cal S}^\star \left(\mathfrak{g}_1\right)$ is a
direct sum of terms of the form

\beqa
\label{Pn}
{\cal S}^k\left( {\cal S}^2 \left({\mathbb C}^n \right) \right)
\otimes {\cal S}^{p-k}\left( \Lambda^{n-2} \left({\mathbb C}^n\right) \right)
\eeqa

\noindent
This representation is in general reducible but we do not know 
of a simple  general formula for the dimension of $\mathfrak{sl}(n)$
invariants.  
\subsection{Trivial and induced $F-$Lie algebras}
In this section $F-$Lie algebras associated to Lie superalgebras
will be constructed explicitly. To fix our notations, consider
$\mathfrak{g}= \mathfrak{g}_0 \oplus \mathfrak{g}_1$ a classical Lie
superalgebra. Let $J_a, 1 \le a \le \mathrm {dim~ }\mathfrak{g}_0$
be a basis of $\mathfrak{g}_0$  and $F_\alpha,
1 \le \alpha \le \mathrm {dim~ }\mathfrak{g}_1$ be a basis of 
$\mathfrak{g}_1$. The structure constants of $\mathfrak{g}$ are given
by 

\beqa
\label{eq:lie}
\left[ J_a, J_b \right]&=& f_{ab}^{~~~c} J_c \nonumber  \\
\left[ J_a, F_\alpha \right]&=& (R_{a})_{\alpha}^{~~\beta} F_\beta,  \\
\left\{F_\alpha, F_\beta\right\}& =& E_{\alpha \beta} = 
S_{\alpha \beta}^{a} J_a
\nonumber
\eeqa  

\noindent
The structure constants  are given {\it e. g.} in \cite{fss} for particular
choices of bases.

The first type of $F-$Lie algebras associated to $\mathfrak{g}$ will
be called trivial and  are constructed as follows:

\begin{theo}\label{trivial}
Let $\mathfrak{g}=\mathfrak{g}_0 \oplus  \mathfrak{g}_1$ be a Lie
superalgebra and let $F \ge 1$ be an integer. Then
$S=\mathfrak{g}_0 \oplus \left( \mathfrak{g}_1 \oplus 
\mathbb C \right)$ (with $ \mathbb C$ the trivial representation
of $\mathfrak{g}_0$) can be given the structure of an  $F-$Lie algebra.
\end{theo}

\noindent
{\bf Proof:}.
The proof is analogous to the proof of \ref{lie-trivial}.
{\bf QED}

The second type of $F-$Lie algebras associated to $\mathfrak{g}$ are those
induced from $\mathfrak{g}$ and symmetric forms on $\mathfrak{g}_1$.
Let $\mathfrak{g}= \mathfrak{g}_0 \oplus \mathfrak{g}_1$ be one of the
classical Lie  superalgebras in the statement of \ref{super} and let $g$ be
a $\mathfrak{g}_0$ invariant symmetric form of order $m$ on
$\mathfrak{g}_1$. The bracket of the associated $F-$Lie algebra of order $m+2$
in the above basis is given by (\ref{eq:tensor})

\beqa
\label{super-F}
\left\{F_{\alpha_1}, \cdots, F_{\alpha_{m+2}} \right\}=
\frac{1}{m!}
\sum \limits_{ i <  j}
g_{\alpha_1 \cdots \alpha_{i-1} \alpha_{i+1} \cdots   
\alpha_{j-1} \alpha_{j+1} \cdots \alpha_{m+2}} E_{\alpha_i \alpha_j}
\eeqa

%\begin{rema}
%By remark \ref{irred-flie},
%these induced  $F-$Lie algebras admit a
%$\bb Z_2-$graded structure instead of a $\bb Z_{2F}$ one.
%\end{rema} 

\begin{ex} \label{sl}
We denote by $S$ the $F-$Lie algebra of order $4$ induced from the
Lie superalgebra

$$A(m-1,n-1)  =
\Big(
\mathfrak{sl}(m) \oplus \mathfrak{sl}(n) \oplus \mathfrak{gl}(1)
\Big) \oplus \left(
{\mathbb C}^{m } \otimes {\mathbb C}^{n \star  }  \otimes {\mathbb C} 
\right) \oplus \left(
{\mathbb C}^{m } \otimes {\mathbb C}^{n \star }  \otimes {\mathbb C}
\right)^\star,$$ 

\noindent
and the tautological quadratic form on
$\left(
{\mathbb C}^{m } \otimes {\mathbb C}^{n \star }  \otimes {\mathbb C} 
\right) \oplus \left(
{\mathbb C}^{m } \otimes {\mathbb C}^{n \star }  \otimes {\mathbb C}
\right)^\star$. 
Let $\left\{E_{IJ}\right\}_{\begin{tiny}\begin{array}{l} 
1 \le I \le m \\ 1 \le J \le m
\end{array}\end{tiny}}$ and
$\left\{E_{IJ}\right\}_{\begin{tiny}\begin{array}{l} 
m+1 \le I \le m+ n \\ m+ 1 \le J \le m+ n
\end{array}\end{tiny}}$ be the standard bases of $\mathfrak{gl}(m)$ and
$\mathfrak{gl}(n)$  respectively.
Let $\left\{F_{IJ}\right\}_{\begin{tiny}\begin{array}{l} 
1 \le I \le m \\ m+1 \le J \le m+n
\end{array}\end{tiny}}$ and
$\left\{F_{IJ}\right\}_{\begin{tiny}\begin{array}{l} 
m+1 \le I \le m+ n \\ 1 \le J \le m
\end{array}\end{tiny}}$
be bases of $(\overline{\mathbf{m}}, \mathbf{n})^+,$ and 
$(\mathbf{m}, \overline{\mathbf{n}})^-$  respectively.

%Then $h=-\frac{1}{m-n}
%(n E_{11} + \cdots +n E_{mm}+ m E_{m+1,m+1}+ \cdots 
%m F_{n+m,n+m}) \in \mathfrak{u}(1)$,
%$E_{IJ}-\Big(\frac{1}{m-n} Z +\frac{1}{m} h \Big) \delta_{IJ} 
%\in \mathfrak{sl}(m), 
%1 \le I,J \le m$
%and
%$E_{IJ}+\Big(\frac{1}{m-n} Z + \frac{1}{n} h \Big) \delta_{IJ} \in 
%\mathfrak{sl}(n), m+1 \le I,J \le m+m$ form a basis of the bosonic part.
%Here 
%$E_{IJ}$ with $ 1 \le I,J \le m$ or $m+1 \le I,J \le n+m$. Then
%the bosonic generators are
%$h=-\frac{1}{m-n}
%(n E_{11} + \cdots +n E_{mm}+ m E_{m+1,m+1}+ \cdots 
%m F_{n+m,n+m}) \in \mathfrak{u}(1)$. Then introduce 
%$Z= E_{11} + \cdots  + F_{n+m,n+m}$. The other bosonic generators write
%$E_{IJ}-\Big(\frac{1}{m-n} Z +\frac{1}{m} h \Big) \delta_{IJ} 
%\in \mathfrak{sl}(m), 
%1 \le I,J \le m$
%and
%$E_{IJ}+\Big(\frac{1}{m-n} Z + \frac{1}{n} h \Big) \delta_{IJ} \in 
%\mathfrak{sl}(n), m+1 \le I,J \le m+m$.
%In the case of  unitary superalgebras, the invariant tensor is given
%by $\delta_{IJ}$.

\noindent
Then the four brackets of $S$ have the following simple form: 

\beqa
\label{eq:unitary}
\left\{F_{I_1 J_1},F_{I_2 J_2},F_{I_3 J_3},F_{I_4 J_4} \right\}& =& 
\delta_{I_1 I_2} \delta_{J_1 J_2} 
\left(\delta_{I_3 J_4} E_{J_3 I_4} + \delta_{J_3 I_4} E_{I_3 J_4} \right)
\nonumber \\
&+&
\delta_{I_1 I_3} \delta_{J_1 J_3} 
\left(\delta_{I_2 J_4} E_{J_2 I_4} + \delta_{J_2 I_4} E_{I_2 J_4} \right)
\nonumber \\
&+&
\delta_{I_1 I_4} \delta_{J_1 J_4} 
\left(\delta_{I_2 J_3} E_{J_2 I_3} + \delta_{J_2 I_3} E_{I_2 J_3} \right)
\\
&+&
\delta_{I_2 I_3} \delta_{J_2 J_3} 
\left(\delta_{I_1 J_4} E_{J_1 I_4} + \delta_{J_1 I_4} E_{I_1 J_4} \right)
\nonumber \\
&+&
\delta_{I_2 I_4} \delta_{J_2 J_4} 
\left(\delta_{I_1 J_3} E_{J_1 I_3} + \delta_{J_1 I_3} E_{I_1 J_3} \right)
\nonumber \\
&+&
\delta_{I_3 I_4} \delta_{J_3 J_4} 
\left(\delta_{I_1 J_2} E_{J_1 I_2} + \delta_{J_1 I_2} E_{I_1 J_2} \right).
\nonumber
\eeqa

\noindent
The fact that the R.H.S is in $\mathfrak{sl}(m) \oplus
\mathfrak{sl}(n)\oplus \mathfrak{gl}(1)$ is a consequence of theorem 
\ref{tensor}.
\end{ex}

\begin{ex}\label{osp}
We denote by $S$ the $F-$Lie algebra of order $4$ induced from the
Lie superalgebra 

$$\mathfrak{osp}(2|2m)= \left(\mathfrak{so}(2) \oplus \mathfrak{sp}(2m)\right)
\oplus {\mathbb C}^2 \otimes {\mathbb C}^{2m},$$

\noindent
and the quadratic form $g=\varepsilon \otimes \Omega$, where 
$\varepsilon$ is the invariant symplectic form on   ${\mathbb C}^2$
and $\Omega$ the   invariant symplectic form on   ${\mathbb C}^{2m}$.
Let $\left\{S_{\alpha \beta} = S_{\beta \alpha }
\right\}_{\begin{tiny}\begin{array}{l} 
1 \le \alpha \le 2m \\ 1 \le \beta \le 2 m
\end{array}\end{tiny}}$  be a basis of $\mathfrak{sp}(2m)$ and 
$\left\{h \right\}$ be a basis of  $\mathfrak{so}(2)$.
Let $\left\{F_{q \alpha}\right\}_{\begin{tiny}\begin{array}{l} 
q=\pm 1\\ 1 \le \alpha \le 2m
\end{array}\end{tiny}}$
be a basis of ${\mathbb C}^2 \otimes {\mathbb C}^{2m}$.
Then the four brackets of $S$ take the following form

\beqa
\label{eq:flie-orth}
&&\hskip 3.truecm 
\left\{F_{q_1  \alpha_1}, F_{q_2  \alpha_2}, F_{q_3 \alpha_3},
F_{q_4  \alpha_4} \right\}=  \\
&&\varepsilon_{q_1 q_2}  \Omega_{\alpha_1  \alpha_2}
\left(\delta_{q_3 + q_4}  S_{\alpha_3 \alpha_4} + 
 \varepsilon_{q_3 + q_4} \Omega_{\alpha_3 \alpha_4} h
\right) 
+
\varepsilon_{q_1 q_3}  \Omega_{\alpha_1  \alpha_3}
\left(\delta_{q_2 + q_4}  S_{\alpha_2 \alpha_4} + 
\varepsilon_{q_2 +q_4}   \Omega_{\alpha_2 \alpha_4} h
\right) \nonumber \\
&+ &
\varepsilon_{q_1 q_4}  \Omega_{\alpha_1  \alpha_4}
\left(\delta_{q_2 + q_3}  S_{\alpha_2 \alpha_3} 
+  \varepsilon_{q_2 + q_3} \Omega_{\alpha_2 \alpha_3} h
\right)
 + 
\varepsilon_{q_2 q_3}  \Omega_{\alpha_2  \alpha_3}
\left( \delta_{q_1 + q_4}  S_{\alpha_1 \alpha_4} + 
 \varepsilon_{q_1 + q_4} \Omega_{\alpha_1 \alpha_4} h
\right) \nonumber \\
&+&
\varepsilon_{q_2 q_4}  \Omega_{\alpha_2  \alpha_4}
\left(\delta_{q_1 + q_3}  S_{\alpha_1 \alpha_3} + 
 \varepsilon_{q_1 + q_3} \Omega_{\alpha_1 \alpha_3} h
\right) 
+
\varepsilon_{q_3 q_4}  \Omega_{\alpha_3  \alpha_4}
\left(\delta_{q_1 + q_2}  S_{\alpha_1 \alpha_2} + 
 \varepsilon_{q_1 + q_2} \Omega_{\alpha_1 \alpha_2} h
\right). \nonumber
\eeqa

\end{ex}

Other extensions of Lie superalgebras have been considered 
in the literature. For instance
 extensions of the orthosymplectic superalgebra $\mathfrak{osp}(1|4)$
or the unitary $\mathfrak{sl}(4|1)$ were constructed 
by means of parafermions and
parabosons \cite{lr}. The first example of an $F-$Lie algebra was considered in
\cite{fvir, fvir2} as a possible extension of the Virasoro algebra. In
\cite{ayu} an example of a ``trivial'' $F-$Lie algebra,
related to the superalgebra $\mathfrak{osp}(1|2)$ was constructed.

\begin{rema}
By repeated application of theorem \ref{tensor}  one construct
$F-$Lie algebras of higher and higher order.
\end{rema}

\section{Finite-dimensional FSUSY extensions of the Poincar\'e algebra}
\renewcommand{\theequation}{5.\arabic{equation}}
\setcounter{equation}{0} 

It is well known that supersymmetric extensions of the Poincar\'e
algebra can be obtained by  In\"on\"u-Wigner contraction
of certain Lie superalgebras. In fact, one can also obtain FSUSY extensions
of the Poincar\'e algebra  by In\"on\"u-Wigner contraction of 
certain $F-$Lie algebras  as we now show  with two
examples.

For the first example, we let
$S_3= \mathfrak{sp}(4) \oplus \mathrm{ad} \   \mathfrak{sp}(4)$
be the real $F-$lie algebra of order three (see Remark \ref{induced-graded})
induced from the real graded $1-$Lie
algebra $S_1= \mathfrak{sp}(4) \oplus \mathrm{ad}  \  \mathfrak{sp}(4)$
(see Example \ref{graded})
and the Killing form on $\mathrm {ad} \ {\mathfrak sp}(4)$.
Using vector indices of $\mathfrak{so}(1,3)$ coming from the inclusion
$\mathfrak{so}(1,3) \subset \mathfrak{so}(2,3) 
\cong \mathfrak{sp}(4)$, the bosonic part of
$S_3$ is generated by $M_{\mu \nu}, M_{\mu 4}$, with 
$\mu, \nu =0,1,2,3$ and the graded part by $J_{\mu \nu}, J_{4 \mu}$.
Letting $\lambda \to 0$
after  the In\"on\"u-Wigner contraction,

\beqa
\begin{array}{ll}
M_{\mu \nu} \to L_{\mu \nu},& M_{\mu 4} \to \frac{1}{\lambda} P_\mu 
\\
J_{\mu \nu} \to \frac{1}{\sqrt[3]{\lambda}} Q_{\mu \nu},& 
J_{4 \mu} \to \frac{1}{\sqrt[3]{\lambda}} Q_{\mu},
\end{array}
\eeqa

\noindent   
one sees that  
$L_{\mu \nu }$ and $P_\mu$ generate the $(1+3)D$ Poincar\'e
algebra and that $Q_{\mu \nu}, Q_\mu$ are the fractional supercharges
 in  respectively  the adjoint and vector representations of
$\mathfrak{so}(1,3)$.
This $F-$Lie algebra of order three is therefore a non-trivial
extension of the Poincar\'e algebra where translations are cubes
of more fundamental generators. The subspace  generated by 
$L_{\mu \nu}, P_\mu, Q_\mu$ is also an $F-$Lie algebra of order three
extending the Poincar\'e algebra  in which
the trilinear symmetric brackets have the simple form: 

\beqa
\left\{Q_\mu, Q_\nu, Q_\rho \right \}=
\eta_{\mu \nu} P_\rho +  \eta_{\mu \rho} P_\nu + \eta_{\rho \nu} P_\mu,
\eeqa

\noindent
where $\eta_{\mu \nu}$ is  the Minkowski metric.
This algebra should be compared to the algebra  recently  obtained in
a different context, where a ``trilinear'' extension of the Poincar\'e
algebra involving   ``supercharges'' in the
 vector representation was constructed \cite{wt}.

For the second example, we let 
$S_4=\left(  \mathfrak{so}(2)\oplus \mathfrak{sp}(4) \right)
\oplus \underline{{\mathbf 2}} \otimes \underline{{\mathbf 4}}$ 
be the real $F-$Lie algebra of order four induced from $\mathfrak{osp}(2|4)$
and the symmetric form $\varepsilon \otimes  \Omega$ , where $\Omega$ is
the symplectic form on $\underline{{\mathbf 4}}$ and $\varepsilon$ the
antisymmetric two-form on $\underline{{\mathbf 2}}$.
Using spinor indices coming from $\mathfrak{sl}(2,\bb C) 
\cong \mathfrak{so}(1,3) \subset \mathfrak{so}(2,3)$ the bosonic part 
is generated by
  $E_{\alpha \beta}, 
E_{\dot \alpha \dot \beta},
E_{\dot \alpha \beta}$ and the fermionic  part by $F_\alpha^\pm,
\bar  F_{\dot \alpha}^\pm, \alpha, \beta =1,2 $ and $\dot \alpha,
 \dot \beta= \dot 1,
\dot 2$.
Letting $\lambda \to 0$ after the In\"on\"u-Wigner contraction
\beqa
\label{eq:iw}
%\lambda E_{\alpha \beta} &\to& \lambda J_{\alpha \beta} + 
% Z_{\alpha \beta} \nonumber \\
%\lambda E_{\dot \alpha \dot \beta} &\to& \lambda J_{\dot \alpha \dot \beta} + 
% Z_{\dot \alpha \dot \beta} \nonumber  \\
%\lambda E_{\alpha \dot \alpha } &\to&  P_{\alpha \dot \alpha} \\
% \lambda h &\to& Z \nonumber \\
%\sqrt[4]{\lambda}F_{\alpha}^\pm &\to&  Q_\alpha^\pm \nonumber  \\
%\sqrt[4]{\lambda}F_{\dot \alpha}^\pm &\to&  Q_{\dot \alpha}^\pm \nonumber
\begin{array}{llll}
E_{\alpha \beta} \to   L_{\alpha \beta}& 
E_{\dot \alpha \dot \beta} \to L_{\dot \alpha \dot \beta}
&
E_{\alpha \dot \alpha } \to  \frac{1}{\lambda} P_{\alpha \dot \alpha} &
 h \to \frac{1}{\lambda} Z \cr
&F_{\alpha}^\pm \to \frac{1}{\sqrt[4]{\lambda}} Q_\alpha^\pm 
&\bar F_{\dot \alpha}^\pm \to \frac{1}{\sqrt[4]{\lambda}} 
\overline{Q}_{\dot \alpha}^\pm,    \cr
\end{array}
\eeqa 

\noindent
one sees that  $L_{\alpha \beta},L_{\dot \alpha \dot \beta}$
and $ P_{\alpha \dot \alpha}$ generate the $(1+3)D$ Poincar\'e algebra,
 that $Z$ is  central  and that $Q_\alpha^\pm, 
\overline{Q}_{\dot \alpha}^\pm$ are the fractional-supercharges in the spinor
representations of $\mathfrak{so}(1,3)$.
This $F-$Lie algebra of order four is therefore a non-trivial
extension of the Poincar\'e algebra where translations are fourth powers
of more fundamental generators. The  four bracket can be  expressed 
simply if we introduce the following notation:
$\sigma_{\alpha \dot \alpha}^\mu, \overline{\sigma}^{\mu \dot \alpha \alpha}$
are  the Dirac matrices,
$\sigma^{\mu \nu}_{\alpha \beta}$,
$\bar \sigma^{\mu \nu}_{\dot \alpha \dot \beta}$
 and $P_\mu$ are the Poincar\'e
generators  (for details {\it e.g.}
\cite{wb}). One then has:

\beqa
\label{eq:poincare}
&& \hskip -1.truecm\left\{Q_{\alpha_1}^{q_1},Q_{\alpha_2}^{q_2}, 
Q_{\alpha_3}^{q_3}, 
Q_{\alpha_4}^{q_4}
\right\}= \nonumber \\
&& 2  \varepsilon^{q_1 q_2} \varepsilon^{q_3 q_4}
 \varepsilon_{\alpha_1 \alpha_2} \varepsilon_{\alpha_3 \alpha_4} Z  
 + 2 \varepsilon^{q_1 q_4} \varepsilon^{q_2 q_3}
 \varepsilon_{\alpha_1 \alpha_4} \varepsilon_{\alpha_2 \alpha_3} Z 
 + 2 \varepsilon^{q_1 q_3} \varepsilon^{q_2 q_4}
  \varepsilon_{\alpha_1 \alpha_3}\varepsilon_{\alpha_2 \alpha_4} Z
\nonumber \\
 \\
&&\hskip-.85truecm\left\{Q_{\alpha_1}^{q_1},Q_{\alpha_2}^{q_2}, 
Q_{\alpha_3}^{q_3}, 
\overline{Q}_{\dot \alpha_4}^{q_4}
\right\}= \delta^{q_1 +q_4} \varepsilon^{q_2 q_3} 
\varepsilon_{\alpha_2 \alpha_3} \sigma^\mu_{\alpha_1 \dot \alpha_4} P_\mu
\nonumber \\
&&\hskip 2.95truecm+ \delta^{q_2 +q_4} \varepsilon^{q_1 q_3} 
\varepsilon_{\alpha_1 \alpha_3} \sigma^\mu_{\alpha_2 \dot \alpha_4} P_\mu
\nonumber \\
&&\hskip 2.95truecm+ \delta^{q_3 +q_4} \varepsilon^{q_1 q_2} 
\varepsilon_{\alpha_1 \alpha_2} \sigma^\mu_{\alpha_3 \dot \alpha_4} P_\mu
\nonumber \\
\nonumber \\
&&\hskip -.75truecm \left\{Q_{\alpha_1}^{q_1},Q_{\alpha_2}^{q_2}, 
\overline{Q}_{\dot \alpha_3}^{q_3}, 
\overline{Q}_{\dot \alpha_4}^{q_4}
\right\}= 0,
\nonumber
\eeqa

\noindent
together with similar  relations involving 
$\left\{Q_{\alpha_1}^{q_1},\overline{Q}_{\dot \alpha_2}^{q_2}, 
\overline{Q}_{\dot \alpha_3}^{q_3}, \overline{Q}_{\dot \alpha_4}^{q_4}
\right\}$ and
$\left\{\overline{Q}_{\dot \alpha_1}^{q_1},\overline{Q}_{\dot \alpha_2}^{q_2}, 
\overline{Q}_{\dot \alpha_3}^{q_3}, \overline{Q}_{\dot \alpha_4}^{q_4}
\right\}$. \\

Analogous constructions   lead to FSUSY extensions of the Poincar\'e
algebra in   any space-time dimensions.

\section{Simple $F-$Lie algebras}
\renewcommand{\theequation}{6.\arabic{equation}} 
 \setcounter{equation}{0}    
%As a consequence of the two previous sections one observes that starting
%with a Lie algebra or a Lie superalgebra, one is able to construct  four
%families of $F-$Lie algebras. 
%So one may wonder of
%the relation between these structures. Under what assumptions if we
%have an $F-$Lie algebra of a given order, we automatically have an
%$F-$Lie algebra of another order. In fact it turns out that these structures
%are totally independent (even between $F-$Lie algebras and Lie
%(super)algebras) but it might happen that some of the representations of 
%$F-$Lie algebras of different order do coincide as we will see in the next 
%section.

%\subsection{Ideals}

By analogy with the case of  Lie (super)algebras we define  
ideals and  the notion of simplicity for $F-$Lie algebras. 

\begin{defi} \label{ideal}
Let $S={\cal B} \oplus {\cal F}$ be an $F-$Lie algebra, 
 or a graded $1-$Lie algebra. Then
$\mathfrak{I}={\cal B}^\prime \oplus {\cal F}^\prime$
 is an ideal of $S$ if and only if 

(i) $\forall f^\prime_1 \in {\cal F}^\prime,\forall f_2,\cdots, f_F \in 
{\cal F}:$
$\left\{f_1^\prime, f_2,\cdots, f_F \right\} \in {\cal B}^\prime;$

(ii) ${\cal B}^\prime$ is an ideal of ${\cal B}$ (
$\forall b^\prime\in  {\cal B}^\prime,\forall b \in {\cal B}, 
[b, b^\prime] \in {\cal B}^\prime$);

(iii) $\forall b \in {\cal B},\forall f^\prime \in {\cal F}^\prime$
$ [b,f^\prime] \in {\cal F}^\prime;$

(iv) $\forall b^\prime \in {\cal B}^\prime,\forall f \in {\cal F}$
$ [b^\prime,f] \in {\cal F}^\prime.$

\end{defi}

\begin{rema}
For a graded $1-$Lie algebra $S={\cal B} \oplus {\cal F}$, denoting $\mu$
 the map from
${\cal F}$ to ${\cal B}$, the property (i) of \ref{ideal} becomes 
$\mathrm {Im} \mu \subset {\cal B}^\prime$.
\end{rema} 

\begin{rema}\label{surjective}
By \ref{ideal}, $\mathrm{ Im} \mu \oplus {\cal F}$ is an ideal of 
$S$ ($\mu$ denotes the 
${\cal B}-$equivariant map from
${\cal S}^F\left({\cal F}\right)$ \-$ \longrightarrow {\cal B}$).
\end{rema}

\begin{rema}
In the case of Lie algebras and Lie superalgebras, this  is the usual
definition.  In the case of a graded $1-$Lie algebra 
$S={\cal B} \oplus {\cal F}$, $S^\prime={\cal B^\prime} \oplus {\cal F^\prime}$
is an ideal if and only if it is a $\mathbb Z_2-$graded ideal for the
natural Lie bracket on $S$ ({\it c.f.} \ref{natural-lie}). 
\end{rema}

%\subsection{Simple $F-$Lie algebras}

\begin{defi}
An $F-$Lie algebra $S$ is said to be simple if and only if its only
ideals  are $S$ and  $\left\{0\right\}$, and 
$\mu :{\cal S}^F\left({\cal F}\right) \to {\cal B}$ is non-zero.
\end{defi}

\begin{rema}
Let $S= {\cal B} \oplus {\cal F}$ be  a graded $1-$Lie algebra such  
that $\mu :{\cal F} \to {\cal B}$
 is non-zero. 
Then,  $S$ is simple if 
 and only if  ${\cal B}$ is a  simple  Lie algebra and 
${\cal F}$ is an irreducible representation of ${\cal B}$. 
\end{rema}

\begin{rema}
If $\mathfrak{g}$ is a  simple Lie algebra, and $S= \mathfrak{g} \oplus
\mathrm{ad~} \mathfrak{g}$ is the graded $1-$Lie algebra of 
Example \ref{graded},
then $S$ is simple as a graded $1-$Lie algebra but is not simple 
as a Lie algebra, with respect to the
natural Lie bracket \ref{natural-lie}.
\end{rema}

\begin{prop} \label{simple-bis}
Let $S= {\cal B} \oplus {\cal F}$ be an $F-$Lie algebra such that  
(i) ${\cal B}$ is semi-simple, (ii) the map 
$\mu :$
${\cal S}^F \left({\cal F}\right) \longrightarrow {\cal B}$ is a surjection
and (iii) no non-zero ideal of ${\cal B}$ has non-zero  fixed points in 
${\cal F}$.  Then:

\begin{enumerate}
\item[(a)] $S$ is simple.
\item[(b)]  The $F-$Lie algebra of order $(F+2)$ induced from  
a ${\cal B}-$equivariant  non-degenerate quadratic form
on ${\cal F}$
(see \ref{tensor}-\ref{tensor-bis}) also satisfies (i) and (ii). 
\end{enumerate}

%${\cal F}$ does not admit a trivial representation for any ideal
%of ${\cal B}$, then $S$ is simple.

\end{prop}

\noindent 
{\bf Proof~:} 
Let $\mathfrak{I}= {\cal B}^\prime \oplus {\cal F}^\prime$ be a non-trivial 
ideal of $S$. Then ${\cal B}^\prime$
is an ideal of ${\cal B}$ and $[{\cal B}^\prime, {\cal F}] \subset 
{\cal F}^\prime$. But if ${\cal F} ={\cal F}^\prime \oplus 
{\cal F}^{\prime \prime}$ as  ${\cal B}^\prime-$modules then 
$[{\cal B}^\prime, {\cal F}^{\prime \prime}]=0$ and therefore, 
${\cal F}^{\prime \prime}= \{0\}$ since by hypothesis ${\cal B}^\prime$
does not admit non-zero fixed points. This proves (a).\\ 
To prove (b). it is enough to prove that the induced $(F+2)-$bracket
is surjective. 
Since the $F-$bracket  $\mu : {\cal S}^{F}\left({\cal F}\right)
\to {\cal B}$ is surjective, by diagonalising the quadratic
form, it is easy to see
that the $(F+2)-$bracket  (\ref{eq:tensor})
is also surjective.  
{\bf QED}

\begin{rema}
If $\mathfrak{g}$ is a simple Lie algebra,
the graded $1-$Lie algebras  $\mathfrak{g} \oplus {\mathrm {ad} 
} \mathfrak{g}$ satisfies (i),  (ii) and (iii) above. As one can check, the 
Lie superalgebras in the list \ref{super} also satisfy (i), (ii) and (iii).
Thus the induced $F-$Lie algebras associated to non-degenerate quadratic
forms and these graded $1-$Lie algebras or Lie superalgebras are
always  simple. 
\end{rema}

%\subsection{Non-simple $F-$Lie algebras}

The trivial $F-$Lie algebras associated to graded $1-$Lie algebras 
or Lie superalgebras \ref{lie-trivial}-\ref{trivial} are not simple
since in both cases ${\mathfrak{g}_0} \oplus {\mathfrak{g}_1 }$ is an ideal of 
$S$.
In particular, when $F=2$,  the trivial Lie superalgebras  
associated to graded $1-$Lie algebras are not simple.
The direct sum of two simple $F-$Lie algebras of the same order  
is clearly not simple.
These two kinds of examples of non-simple $F-$Lie algebras indicate
 that probably,
as for  Lie superalgebras, there are different inequivalent ways
to define semi-simple Lie $F-$Lie algebras. 

\section{Representations}
\renewcommand{\theequation}{7.\arabic{equation}}
\setcounter{equation}{0}      
\begin{defi}
A representation of an $F-$Lie algebra 
 $S$  is  a linear map
$\rho : ~ S \to \mathrm{End}(H)$, 
and a 
automorphism $\hat \varepsilon$ such that $ \hat \varepsilon^F=1$ 
which satisfy 

\beqa      
\label{eq:rep}
\begin{array}{ll}
\mathrm{a)}& \rho\left(\left[x,y\right]\right)= \rho(x) \rho(y)-  
\rho(y)\rho(x) \cr
\mathrm{b)}& \rho \left\{a_1.\cdots,a_F\right\}=
 \sum \limits_{\sigma \in S_F}
\rho\left(a_{\sigma(1)}\right) \cdots \rho\left(a_{\sigma(F)}\right) \cr
\mathrm{d)}& \hat \varepsilon \rho\left(s\right) \hat \varepsilon^{-1} =
\rho\left(\varepsilon\left(s\right)\right)
\end{array}
\eeqa

\noindent
($S_F$ being the group of permutations of $F$ elements).
\end{defi}

As  a consequence of these properties,  
since the eigenvalues of $\hat \varepsilon$ are $\mathrm{F}^{\mathrm{th}}-$
roots of unity, we have  the following decomposition 

$$H= \bigoplus \limits_{k=0}^{F-1} H_k,$$

\noindent
where $H_k=\left\{\left|h\right> \in H ~:~ 
\hat \varepsilon\left|h\right>=q^k \left|h\right> \right\}$.
The operator $N \in \mathrm{End}(H)$
defined by $N\left|h\right>=k
\left| h \right>$ if $\left|h\right> \in H_k$
is the  ``number operator'' (obviously $q^N=\hat \varepsilon$).
Since $\hat \varepsilon \rho(b)= \rho(b) \hat \varepsilon, \forall b \in {\cal B}$
each $H_k$ provides a representation of the Lie algebra ${\cal B}$. 
Furthermore, for $a \in {\cal A}_\ell$,
 $\hat \varepsilon \rho(a)=q^\ell \rho(a) \hat \varepsilon$ and so
we have 
$\rho(a) .H_k\ \subseteq 
H_{k+\ell ({\mathrm{mod~} F)}}$  \\

\begin{ex} \label{so-sp}
Let $X,Y,Z$ be   $n\times n$ (resp.  $2n \times 2n$) matrices in 
$\mathfrak{so}(n)$
(resp.  $\mathfrak{sp}(2n)$). Then, it is
easy to see that $\{X,Y,Z\}$ is also in $\mathfrak{so}(n)$ (resp.
$\mathfrak{sp}(2n)$).
Consequently, $S=\mathfrak{so}(n)\oplus \mathfrak{so}(n)$
(resp. $S=\mathfrak{sp}(2n) \oplus \mathfrak{sp}(2n)$), is an $F-$Lie
algebra of order $3$ (the only non-trivial point to be checked is the Jacobi 
identity 
(J4) in Definition \ref{f-lie}). 
A similar property is true  for any odd number of matrices.
We will  calculate the structure constants in the case of 
$\mathfrak{so}(n)$, the calculation for $\mathfrak{sp}(2n)$ being
analogous. 
If $X_a,  1 \le a \le {\mathrm {dim}} \  \mathfrak{so}(n)$ is a basis of 
$\mathfrak{so}(n)$, then the $3-$bracket of $S$ is given by

\beqa
\label{eq:so-sp}
\left\{X_a,X_b,X_c \right\}= k_{abc}^{\ \ \ \ d} X_d.
\eeqa

\noindent
Writing $\left\{X_a,X_b,X_c,X_d\right\}= \Big(
\left\{X_a,X_b,X_c\right\}X_d +
\left\{X_a,X_b,X_d\right\}X_c +
\left\{X_a,X_c,X_d\right\}X_b +
\left\{X_b,X_c,X_d\right\}X_a \Big)$
and taking the trace using  (\ref{eq:so-sp}), we get
$4 k_{abc}^{\ \ \ \ d} tr(X_d X_e) = \mathrm{Tr}\left(\left\{X_a,X_b,X_c,X_e 
\right\} \right)$. Since the trace defines a metric on $\mathfrak{so}(n)$
this gives $k_{abc}^{\ \ \ \ d}= \frac{1}{4} {\mathrm Tr }
\left\{X_a,X_b,X_c,X_d\right\} g^{de}$. 

This $F-$Lie algebra of order 
three is {\it not } induced from the graded $1-$Lie algebra 
$\mathfrak{so}(n) \oplus \mathfrak{so}(n)$ and the Killing form:
if this where the case we would have 
$\left\{X_a,X_b,X_c\right\}= \mathrm{Tr}\left(X_a X_b \right) X_c +
 \mathrm{Tr}\left(X_a X_c \right) X_b + \mathrm{Tr}\left(X_b X_c \right) X_a$
which is clearly false if $a=b=c$. 
However, by proposition \ref{simple-bis} $S$  is simple.

We can construct a representation of $S$ in ${\mathbb C}^n \otimes
{\mathbb C}^3$ as follows: define $\rho: S \to {\mathrm End} 
\left({\mathbb C}^n \otimes {\mathbb C}^3\right)$ by 

\beqa 
\rho(X) =
\left\{ \begin{array}{ll}
 X \otimes \mathbf {Id} & {~if~} X {~is~in ~the ~first~} 
\mathfrak{so}(n) \cr
 X \otimes Q& {~if~} X {~is~in ~the ~second~}
\mathfrak{so}(n), 
\end{array}
\right.
\eeqa

\noindent    
where $Q: {\mathbb C}^3 \to {\mathbb C}^3$ is any linear map 
whose minimal polynomial is $\lambda^3-1$
({\it i.e.}, $Q^3=\mathbf {Id}$ and $Q$ has three distinct eigenvalues).

Related results were obtained for $\mathfrak{so}(n)$ and $\mathfrak{sp}(2n)$
in \cite{multi-lie}. 
\end{ex}

\begin{ex}\label{sl-n}
Let $X,Y,Z$ be  three $n\times n$ matrices in $\mathfrak{u}(n)$. Then, it is
easy to see that $\{X,Y,Z\}$ is also in $\mathfrak{u}(n)$.
As in the previous example, this simple observation enables us
to give $\mathfrak{u}(n) \oplus \mathfrak{u}(n)$ or 
$\mathfrak{u}(n) \oplus \mathfrak{su}(n)$ the 
 structure of an $F-$Lie algebra of order $3$.
\end{ex}

\begin{ex} \label{mat-sl}
Let $A(m-1,n-1)$, $n \ne m$  be the Lie superalgebra of $(n+m)\times(n+m)$
matrices \cite{fk, fss}, 

$$M=\begin{pmatrix}
E_{mm}& F_{mn} \cr
F_{nm} &E_{nn}
\end{pmatrix},$$

\noindent
of supertrace zero ({\it i.e.}, $\mathrm{sTr} M=\mathrm {tr}E_{mm}- 
\mathrm {tr}E_{nn}=0$).

If $J_{i_1}, \cdots, J_{i_{2F}}$ are arbitrary matrices then

%odd matrices  in $A(n|m)$, then from

\beqa
\label{eq:trace}
\left\{J_{i_1}, \cdots,J_{i_{2F}}\right\}=  
 \sum \limits^F_{\begin{array}{l}a < b =1  \cr  a \ne b \end{array} }
\Big\{\left\{J_{i_a},J_{i_b} \right\}, 
\left\{\hat J_{i_a}, \hat J_{i_b} ,J_{i_1}, \cdots,J_{i_{2F}}\right\} \Big\}. 
%\nonumber \\
%&+&
%\left\{\hat J_{i_a}, \hat J_{i_b}J_{i_1}, \cdots,J_{I_{2F}}\right\}
%\left\{J_{i_a},J_{i_b} \right\} \Big)  
\eeqa

\noindent 
%where $\hat J_{i_a}, \hat J_{i_b}$ means that in the bracket of
%order $2F-2$ the terms  $J_{i_a}$ and $J_{i_b}$ are omitted.
Applying this formula to $2F$ odd matrices in $A(m-1,n-1)$ one sees by 
an induction that the supertrace of the
 $2F-$bracket (\ref{eq:trace}) vanishes. Using  the $\mathbb Z_2$ graduation
of $A(m-1,n-1)$ one sees that this bracket belongs to 
the even part of the algebra
and hence  defines the structure of an  $F-$Lie algebra of order $2F$
on the underlying vector space of $A(m-1,n-1)$.
For $F=4$ this is just the 
the $F-$Lie algebra of order 4 induced by the tautological quadratic
form of Example \ref{sl}. 
Indeed, let
$V={\mathbb C}^{\star n} \otimes {\mathbb C}^{ m} \otimes {\mathbb C}$
and let $\mathfrak{g}_0= \mathfrak{sl}(n) \oplus \mathfrak{sl}(m) \oplus
\mathfrak{gl}(1)$. Then, comparing $\mathfrak{gl}(1)$ charges,
we have 
  $\mathrm{Hom}_{\mathfrak{g}_0}
\left({\cal S}^4 \Big(V \oplus V^\star\Big), \mathfrak{g}_0\right)=
\mathrm{Hom}_{\mathfrak{g}_0}
\left({\cal S}^2 (V) \otimes {\cal S}^2 ( V^\star), \mathfrak{g}_0\right)$.
Since,

$${\cal S}^2 \Big(V  \Big) 
\otimes {\cal S}^2 \Big( V^\star\Big) \cong
\Big({\cal S}^2({\mathbb C}^{\star n}) \otimes {\cal S}^2({\mathbb C}^{m})
\oplus \Lambda^2({\mathbb C}^{\star n}) \otimes \Lambda^2({\mathbb C}^{m})
\Big) \otimes
\Big({\cal S}^2({\mathbb C}^{ n}) \otimes {\cal S}^2({\mathbb C}^{\star m})
\oplus \Lambda^2({\mathbb C}^{n}) \otimes \Lambda^2({\mathbb C}^{\star m})
\Big)$$
 and since the representations ${\mathbf 1}$ and $\mathfrak{sl}(n)$ 
occur exactly once in ${\cal S}^2({\mathbb C}^{ n})\otimes 
{\cal S}^2({\mathbb C}^{\star n})$ and not at all in 
${\Lambda}^2({\mathbb C}^{ n})\otimes {\Lambda}^2({\mathbb C}^{\star n})$,
we deduce that $\mathrm{Hom}_{\mathfrak{g}_0}
\left({\cal S}^4 \Big(V \oplus V^\star\Big), \mathfrak{g}_0\right)$ is
of dimension one.

By definition, the fundamental $(n+m)\times (n+m)$ matrix representation
of the Lie superalgebra $A(m-1,n-1)$ is also a representation of the $F-$Lie
algebra of order $2F$ constructed above.
In general, this is not true: for instance if $m=2, n=1$, one can check that 
the
$6-$dimensional representation of $A(2,1)$ is not a representation of  
the associated $F-$Lie algebra of order $4$.

\end{ex}

\begin{ex} \label{mat-osp}

% This construction is similar
% to the one of $\mathfrak{osp}(m|2n)$ \cite{fk, fss}.
%The invariant tensor   tensor is
%given by  $G=\begin{pmatrix} I&0 \cr 0& \Omega \end{pmatrix}$,
%with $I$ the $ m \times m$ identity matrix and $\Omega$ the $2n \times 2n$
%antisymmetric matrix

% $$\Omega = \begin{pmatrix}0&-1 \cr 1 &0 \end{pmatrix} \otimes 
%\begin{pmatrix}1&\cdots&0 \cr\vdots & \ddots&\vdots \cr 0&\cdots&1 
%\end{pmatrix}.$$

%\noindent
%The symplectic algebra acts on the last $2n$ elements
%and is characterized by    $2n \times 2n$ matrices $S$
%satisfying $S^t=\Omega S \Omega$.  The
%orthogonal one acts on the first $m$ components and are 
%specified by $m \times m $ antisymmetric matrices $J^t=-J$.

Let $S$ be  the set of all matrices of the form 
% $(2n +2) \times (2n +2)$
%matrix, 

\beqa
\label{eq:osp-flie}
M=\begin{pmatrix}
q&0&F_+ \cr
0&-q&F_- \cr 
-\Omega F_-^t&-i \Omega F_+^t&S  
\end{pmatrix},
\eeqa

\noindent
where   $q$ is  a complex  number,             
$F_\pm$ are two $1 \times 2n$ matrices, $\Omega$ is the standard $2n \times 2n$
symplectic form on ${\mathbb C}^{2n}$ 
 and   $S$ is a $2n \times2n$ matrix in $\mathfrak{sp}(2n)$, 
{\it i.e}, $S^t=\Omega S \Omega$. 
Let ${\cal B}=\left\{
\begin{pmatrix}
q&0&0 \cr
0&-q&0 \cr
0&0&S 
\end{pmatrix}, q\in {\mathbb C}, S \in \mathfrak{sp}(2n) \right\} 
\cong  \mathfrak{so}(2) \oplus  \mathfrak{sp}(2n)$
and let ${\cal F}= \left\{
\begin{pmatrix}
0&0&F_+ \cr
0&0&F_- \cr
-\Omega F_-^t&-i \Omega F_+^t&0 
\end{pmatrix}, F_\pm \in {\cal M}_{1,2n}\left({\mathbb C}\right) \right\}$.
If one now takes 

$${\cal F}_{a +}=\begin{pmatrix}
0&0&F_{a+} \cr
0&0&0 \cr
0&-i \Omega F_{a+}^t&0 
\end{pmatrix}, \ \ 
{\cal F}_{a -}=\begin{pmatrix}
0&0&0 \cr
0&0&F_{a-} \cr
-\Omega F_{a-}^t&0&0 
\end{pmatrix}
$$
and ${\cal F}_{a}= {\cal F}_{a +}+ {\cal F}_{a -}$
\noindent
we get 
$\left\{{\cal F}_{a}, {\cal F}_{b} \right\}= 
\begin{pmatrix} \alpha_{ab} & 0 &0 \cr
                0&-i \alpha_{ab} &0 \cr
                0&0&A_{ab}
\end{pmatrix}$, 
where $A_{ab}=-\Omega F_{a-}^t F_{b+} - i \Omega F_{a+}^t F_{b-} 
-\Omega F_{b-}^t F_{a+} - i \Omega F_{b+}^t F_{a-}$ and where 
$\alpha_{ab}= -F_{a+} \Omega F_{b-}^t - F_{b+} \Omega F_{a-}^t.$ This
shows that ${\cal B} \oplus {\cal F}$ is not closed under the 
superbracket. 
From the formula $\left\{{\cal F}_{a_1}, {\cal F}_{a_2},{\cal  F}_{a_3},
 {\cal F}_{a_4} \right\}
=\left\{ \left\{{\cal F}_{ a_1}, {\cal F}_{ a_2} \right\}, 
\left\{{\cal F}_{a_3}, {\cal F}_{a_4} \right\} \right\} +
\left\{ \left\{{\cal F}_{a_1}, {\cal F}_{a_3} \right\}, 
\left\{{\cal F}_{a_2}, {\cal F}_{a_4} \right\} \right\}  +
\left\{ \left\{{\cal F}_{a_1}, {\cal F}_{a_4} \right\}, 
\left\{{\cal F}_{a_2}, {\cal F}_{a_3} \right\} \right\}$, 
observing that $\left\{{\cal F}_{a +},{\cal F}_{b +}\right\}=
\left\{{\cal F}_{a-},{\cal F}_{b-}\right\}=0$ the four bracket
$\left\{{\cal F}_{  a_1 q_1},{\cal  F}_{  a_2 q_2},
{\cal  F}_{  a_3 q_3}, {\cal F}_{ a_4 q_4} \right\}=0$ if 
$q_1+q_2+q_3+q_4 \ne 0$.
We then calculate $4$-brackets  for $q_1=q_2=-q_3=-q_4=1$ and obtain 

\beqa
\label{mat-osp2}
&&  \left\{{\cal F}_{a +},{\cal  F}_{b +},
{\cal  F}_{c -}, {\cal F}_{d -} \right\}= \begin{pmatrix}q&0&0 \cr
                                                         0&-q&0 \cr
                                                        0&0&S \end{pmatrix}, 
\nonumber \\
q&=&2  \left(F_{a +} \Omega F_{c-}^t  \right)
\left( F_{b +} \Omega F_{d-}^t \right) +
2 \left(F_{a +} \Omega F_{d-}^t  \right)
\left( F_{b +} \Omega F_{c-}^t \right)
 \\
S&=& F_{a +} \Omega F_{d-}^t 
\Big(\Omega  F_{c-}^t F_{b+} + \Omega  F_{b +}^t F_{c -}  \Big)
+ F_{a +} \Omega F_{c-}^t
\Big(\Omega  F_{d-}^t F_{b+}  + \Omega  F_{b +}^t F_{d -}  \Big) 
\nonumber \\
&+& F_{b +} \Omega F_{d-}^t 
\Big(\Omega  F_{c-}^t F_{a+} + \Omega  F_{a +}^t F_{c -}  \Big)
+ F_{b +} \Omega F_{c-}^t
\Big(\Omega  F_{d-}^t F_{a+} + \Omega  F_{a +}^t F_{d -}  \Big). \nonumber
\eeqa

\noindent
This shows that ${\cal B} \oplus {\cal F}$
is  an $F-$Lie algebra of order $4$ since  $S^t= \Omega S \Omega$.

In fact, the matrices of ${\cal B}\oplus {\cal F}$ define a 
representation of the $F-$Lie algebra of order $4$
induced from $\mathfrak{osp}(2|2m)$ and $\varepsilon \otimes \Omega$
(see \ref{osp}).
Indeed setting 
$$\overline{{\cal F}}_{a +}=\begin{pmatrix}
0&0&F_{a+} \cr
0&0&0 \cr
0&- \Omega F_{a+}^t&0 
\end{pmatrix}, \ \ 
\overline{{\cal F}}_{a -}=\begin{pmatrix}
0&0&0 \cr
0&0&F_{a-} \cr
-\Omega F_{a-}^t&0&0 
\end{pmatrix},$$
we see that ${\cal B} \oplus \overline{{\cal F}} \cong \mathfrak{osp}(2|2m)$
and that
 $\left\{{\cal F}_{a + }, {\cal F}_{b + },{\cal  F}_{c -},
 {\cal F}_{d -} \right\}=
<\overline{{\cal F}}_{a+},\overline{{\cal F}}_{c-}>
\left\{\overline{{\cal F}}_{b+},\overline{{\cal F}}_{d-}\right\}+
<\overline{{\cal F}}_{b+},\overline{{\cal F}}_{d-}>
\left\{\overline{{\cal F}}_{a+},\overline{{\cal F}}_{c-}\right\}+
<\overline{{\cal F}}_{a+},\overline{{\cal F}}_{d-}>
\left\{\overline{{\cal F}}_{b+},\overline{{\cal F}}_{c-}\right\}+
<\overline{{\cal F}}_{b+},\overline{{\cal F}}_{c-}>
\left\{\overline{{\cal F}}_{a+},\overline{{\cal F}}_{d -}\right\}$

\noindent 
where $<\overline{{\cal F}}_{a+},\overline{{\cal F}}_{c-}>$ denotes the
$\varepsilon \otimes \Omega$ invariant form.

\end{ex}

Given an $F-$Lie algebra $S={\cal B} \oplus {\cal F}$ one can define
the universal enveloping algebra ${\cal U}(S)$ by taking the quotient
of the tensor algebra ${\cal T}(S)$ by the two-sided ideal generated by
(see definition \ref{f-lie})

\begin{eqnarray}
\label{eq:universal}
\left\{
\begin{array}{l}
 \sum\limits_{\sigma \in \Sigma_F}
a_{\sigma(1)} \otimes \cdots \otimes a_{\sigma(F)}
-\left\{a_1,\cdots, a_F \right\}, \\
b_1 \otimes b_2 -b_2 \otimes b_1- \left[b_1,b_2\right], \\
b_1 \otimes a_2 -a_2 \otimes b_1-\left[b_1,a_2\right], \\
\end{array}
\right.
\end{eqnarray}

\noindent
with $a_1, \cdots, a_F \in {\cal A}_1, b_1,b_2 \in {\cal B}$.
It is not necessary to impose the Jacobi identity (J4) since it is
true in ${\cal T}(S)$.

The natural filtration of ${\cal T}(S)$ factors to a filtration of
${\cal U}(S)$ and, denoting the associated graded algebra by 
$\mathrm{gr}({\cal U}(S))$, we conjecture the following:

\begin{enumerate}
 \item[(1)]  $\mathrm{gr}({\cal U}(S))$ is isomorphic to
$ {\cal T}(S)/\bar I$, where $\bar I$ is the two-sided ideal generated by

\begin{eqnarray}
\label{eq:universal2}
\left\{
\begin{array}{l}
 \sum\limits_{\sigma \in \Sigma_F}
a_{\sigma(1)} \otimes \cdots \otimes a_{\sigma(F)},\\
b_1 \otimes b_2 -b_2 \otimes b_1, \\
b_1 \otimes a_2 -a_2 \otimes b_1.
\end{array}
\right. \nonumber 
\end{eqnarray}

\noindent
(This would then imply that
$\mathrm{gr}({\cal U}(S)) \cong S({\cal B}) \otimes \Lambda_F({\cal F})$,
where $S({\cal B})$ is the symmetric algebra on ${\cal B}$ and 
$\Lambda_F({\cal F})$ is the $F-$exterior algebra on ${\cal F}$ \cite{cliff}).

\item[(2)]  The natural
map $\pi: {\cal U}(S) \to \mathrm{gr}({\cal U}(S))$ is a linear isomorphism.
(This would be an analogue of the Poincar\'e-Birkhoff-Witt theorem).
\end{enumerate}

In the usual way, the representations of $S$ 
are in bijective correspondence with the  representations
of the associative algebra ${\cal U}(S)$.
Consequently, if  ${\cal I} \subset {\cal U}(S)$ 
is a   two-sided ideal, then the quotient $ {\cal U}(S)/{\cal I}$ gives 
a representation of $S$. 
It would be  very convenient to have a theory of ``Cartan sub-algebras'', 
``roots'' and ``weights'' for  $S$. However, even for simple Lie
superalgebras this kind of theory only works well  for  basic Lie
superalgebras \cite{fss}. One might expect   $F-$Lie
algebras induced from basic Lie superalgebras to be  amenable 
to this approach. This seems not to be the case. Indeed, recall that  if $S$ is
a basic Lie superalgebra with Borel decomposition $S={\mathfrak h} \oplus 
{\mathfrak n}_+ \oplus {\mathfrak n}_-$ and  
$\lambda \in {\mathfrak h}^\star$  is a dominant weight, then 
${\cal V}_\lambda={\cal U}/{\cal I}_\mu$   
(where ${\cal I}_\lambda$ is the ideal corresponding  to $\lambda$)  
  is (i)  generated by the action of ${\mathfrak n}_+$ on the vacuum 
and (ii)  has a unique quotient ${\cal D_\lambda}$  
on which the action of ${\mathfrak n}_+$  is 
nilpotent and which is therefore finite-dimensional.  
However, if $S_g$ is the $F-$Lie   
algebra induced from $S$ and a  symmetric form $g$, the quotient    
${\cal V}_\lambda^\prime = {\cal U}(S_g)/{\cal I}_\lambda^\prime$   
 is (i)  not  generated by the action of  ${\mathfrak n}_+$ on the  
vacuum and (ii) the  nilpotence of the action of ${\mathfrak n}_+$ 
in a quotient does not guarantee finite-dimensionality.
This means that in finite-dimensional representations of $S$,
as in the examples of section 7, the elements of ${\mathfrak n}_+$
are not only nilpotent but also satisfy  additional relations. \\

\section{Conclusion} 
\setcounter{equation}{0}
The mathematical structure underlying supersymmetry is that of
a Lie superalgebra.
Given the classification of Lie superalgebras,  one can list
the possible supersymmetric extensions of the  Poincar\'e algebra.
These extensions have had  a wide range of applications in physics.

Fractional supersymmetries were first studied in
the early 1990's in relation with 
low dimensional physics $(D \le 3)$ where  fields
which are neither bosonic nor fermionic \cite{anyon} do exist.
 It was understood a few years later that FSUSY 
can be considered  in arbitrary dimensions and the definition of an
$F-$Lie algebra, the underlying mathematical structure,
was given \cite{flie}. However when $F>2$, most of the examples of $F-$Lie
algebras   which have been found since then are of infinite-dimensions.
In this paper, we  show  how one can  construct many finite-dimensional
$F-$Lie algebras starting from   Lie algebras or   Lie superalgebras
equipped with  appropriate symmetric forms. We  define 
a notion of simplicity in this context and show that 
some of our examples are simple.
 Furthermore, we  construct the first finite-dimensional 
FSUSY extensions of the Poincar\'e algebra by In\"on\"u-Wigner
contraction of certain $F-$Lie algebras.

These results can be seen as a first step
in classifying $F-$Lie algebras.

\paragraph*{Acknowledgements}
J.~Lukierski and Ph. Revoy are
gratefully acknowledged for useful discussions and remarks.

%\newpage

\end{document}